\documentclass{kluwer}    
\usepackage{psfig}
\newdisplay{guess}{Conjecture}

\begin{document}                                                                                   
\begin{article}
\begin{opening}         
\title{Spectral Properties of Black Holes in Gamma Rays} 
\author{Sandip Kumar \surname{Chakrabarti}}  
\runningauthor{S.K. Chakrabarti}
\runningtitle{Spectra of Black Holes in Gamma rays}
\institute{S.N. Bose National Centre for Basic Sciences, JD Block, Salt Lake, Kolkata 700098\\
and\\
Centre for Space Physics, Chalantika 43, Garia Station Rd., Kolkata 700084\\ }
\date{August 12th, 2004}

\begin{abstract}
Black holes are the most compact objects in the universe. Therefore, 
matter accreting onto is likely to radiate photons of energy comparable 
to very high gravitational potential energy. We discuss the nature of 
the emitted radiation in X-rays and gamma-rays from black hole candidates.
We present theoretical solutions which comprise both Keplerian and sub-Keplerian
components and suggest that shocks in accretion and outflows 
may play a major role in producing this spectra.
\end{abstract}

\end{opening}           

\noindent  To appear in Astrophysics and Space Science (Proceedings of the Honk Kong Conference,
Edited by Cheng and Romero)

\section{Introduction}  

It is well known that the black holes themselves do not emit any 
radiation, but the matter which accretes on to them does. There are three 
distinct ways one could study the spectral properties: the first is through 
direct observation of the radiation. Here, one has to use separate instruments, either ground based
(IR, optical, UV, very high energy gamma rays) or space based 
(optical, X-rays, gamma-rays). The goal would be to find if there are fast 
variabilities or spectral signatures of high velocity (which may point to massive compact objects 
at the core or some spectral signatures especially predicted for black holes.).  A second
method which is really the other extreme, is to solve equations which govern 
the motion of matter around a black hole and produce the most likely solutions
which explain both the steady and non-steady behaviour of the radiation. 
A third way is intermediate between the other two approaches:
here, one assimilates the gist of the theoretical results
and gives models and fits the observational results with these models 
using several free parameters. (The reverse may also be appropriate sometimes: 
theoreticians may get clues from `models' to select the best solution out of 
many possible ones.) Very often, this third approach requires separate
models for separate objects. 

In the present paper, I will present the theoretical reasonings behind the
most accurate solution of the black hole accretion of today in order to assist the observers 
to more fruitfully plan their observations and perhaps to provide hints to the model 
builders so that they may fine tune their present models accordingly. Historically, 
in the present subject, tentative models came first (such as disks with corona, see, e.g., Galeev, Rosner \& Viana, 1979) 
and as the theoreticians sharpened their tools through viscous transonic solutions 
with heating, cooling, steady and oscillating shocks (Chakrabarti 1990; 1996a), 
both theoretically and through numerical simulations, new models started emerging 
(a good example being Something or the Other Dominated Accretion Flow [SODAF] models; 
dynamical corona models etc.) which converge to the outcome similar 
to these theoretical solutions. Consequently, these models do not throw 
any new insights into the problem.

Because correct theoretical works incorporates proper boundary conditions
(such as entering through the black hole with velocity of light, or hitting the
hard surface of a neutron star with zero radial velocity) the solutions 
got to be correct beyond doubt. However, in certain situations (such as 
inclusion of all the components of the magnetic field self-consistently)
it may not be possible to obtain solutions in the first place. In such cases `patch-up'
works are done (such as putting toroidal field or radial field on the top of 
hydrodynamics solutions) which are often satisfactory.  

\section{The advective disk paradigm}

A cartoon diagram of the flow geometry which comes out of extensive analytical and numerical 
works (see, e.g., Chakrabarti, 1990, Chakrabarti 1993, Chakrabarti, 1996ab) is 
 shown in Fig. 1. The Keplerian disk, because of its low energy, 
resides at the equatorial plane, while matter with lower angular 
momentum flows above and below it (Chakrabarti, 1995; Chakrabarti \& 
Titarchuk, 1995 [CT95]; Ebisawa, Titarchuk \& Chakrabarti 1996; Chakrabarti, 
Titarchuk, Kazanas \& Ebisawa, 1996; Chakrabarti, 1997 [C97]). The wind is predominantly 
produced from the CENBOL area or the CENtrifugal barrier supported 
BOundary Layer of the black hole, which is the post-shock region. A transient shock
may be present just outside the inner sonic point. The inner edge of the Keplerian disk
is terminated at the transition radius as dictated by  the viscosity of the flow.
The wind may be variable because of shock-oscillation and every supersonic puff of wind 
is expected to produce a propagating shock in the jet. It is to be
noted that all the components in this diagram could be variable (see, Chakrabarti \& Nandi, 2000).
Not surprisingly, this paradigm, created almost a decade ago, is now vindicated by 
most of the observations. Indeed, the cartoon diagrams 
(see, e.g., Esin et al, 1998; Zdziarski and Gerlinski, 2004, Novak 2003, Fender et al, 1999) 
are slowly, but surely, converging to those predicted by our exact approach. 
Whereas our exact approach identifies the reason of the movement of the inner
edge of the disk with viscosity and accretion rate, other models merely take these
as facts and proceed fitting with observations assuming such behaviours {\it a priori}.

The advective paradigm, which requires two component advective flows (TCAF)
has been tested for a varied range of observational facts. It is the only solution
which explains QPOs and their variations quite naturally as manifestations of the
shock oscillations (Chakrabarti, Acharyya \& Molteni, 2004 and references therein).
Hence it is worthy to look at the high energy predictions within this paradigm in order to
find an interpretation of recent gamma-ray observations.

\begin{figure}
\vbox{
\vskip 0.0cm
\hskip 0.0cm
\centerline{
\psfig{figure=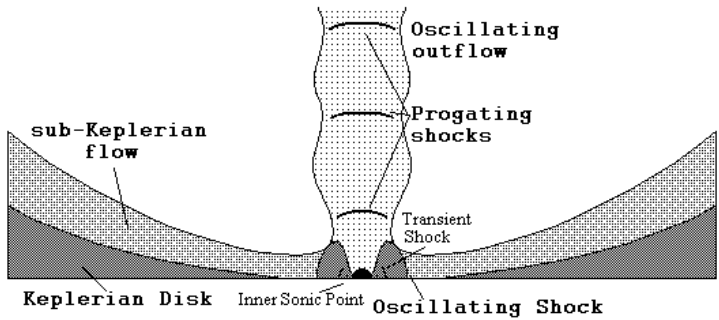,height=5truecm,width=10truecm}}}
\vspace{0.0cm}
\caption[]{Cartoon diagram of the Two Component Advective Flow (TCAF) solution highlighting different
components. The centrifugal pressure dominated boundary layer or CENBOL is the post-shock region
of the combined Keplerian and sub-Keplerian flow. There could be transient shocks close to the
inner sonic point. Propagating shocks are expected to form once in each cycle of 
shock oscillation.}
\end{figure}

\section{Spectral properties at soft and hard X-rays}

Detailed properties of the spectrum coming out of TCAF solution have been discussed in CT95 and C97. The soft radiation 
coming from the Keplerian disk is intercepted by the sub-Keplerian hot flows in the CENBOL region and is re-radiated 
as the hard radiation. Depending on the relative importance of the Keplerian disk rate ${\dot M}_{d}$
in comparison to the sub-Keplerian halo rate ${\dot M}_{h}$, the electrons in the sub-Keplerian 
disk may or may not be cooled down. If not, then the system is in the hard state and if yes, the system
is in the soft state. The hard state is thus dominated by a power-law hard photon component
produced by inverse Comptonization of the soft photons. In the soft state, the 
electrons in CENBOL cooled down and collapses, but due to the inner boundary condition of supersonicity,
the matter accelerates rapidly and up-scatters photons to its energy ($\sim m_e c^2$). Thus,
power-law photons can also be seen due to this, so-called, bulk motion Comptonization (BMC)
(CT95). Occasionally, as we shall show below, photons of much higher energies are seen. But 
that does not imply that the  BMC is not taking place. BMC only asserts what should be the 
{\it minimum} intensity of power-law photons in soft-states as dictated purely by gravitational pull of the 
black hole. Of course, in presence of other physical processes (such as synchrotron radiation,
see, below), very high energy photons should be expected. It is to be noted that CT95 
solution already incorporates all the iterations of re-processing between the Keplerian and
sub-Keplerian components. Therefore, it is no longer necessary to include a `Compton reflection' component
that is usually assumed by some model builders.

Being almost freely falling, the sub-Keplerian flow would bring forth faster changes
to spectral indices than the Keplerian flow which moves in viscous time-scale.  Figure 2 shows
the variation of photon flux and power-law index of Cyg X-1 (upper panel) and for GRS 1758-258 (lower panel)
for more than four years as seen by RXTE. This distinctly different behaviour
of a preferentially wind accretor (Cyg X-1) and the Roche-lobe accretor (GRS 1758-258) 
could be interpreted very clearly when two independent components as predicted by TCAF solution
are used (Smith, Heindl and Swank, 2002). For instance, if there is a reduction in
the accretion rate at the outer edge, 
the sub-Keplerian component at the inner edge will be reduced in free fall time
$t_{ff} \sim r^{3/2}$. Until this information propagates through the Keplerian disk with viscous 
time scale, the source would behave like a soft state, but then, after the viscous time,
the spectrum will harden as the soft photons go down. In Cyg X-1, Cyg X-3 etc. which are known to be wind accretors,
the Keplerian disk is very small and the power-law index basically correlates with the 
photon flux. In 1E 1740.7-2942, GRS 1758-258, and GX 339-4, which are in low-Mass X-ray 
binary systems, the Keplerian disk is big and the delay of tens of days are common (Smith, Heindl and Swank, 2002).
Another very important point made in C97 was that a single component ADAF (Esin et al. 1998) 
model is incapable of producing state transition unless the inner edge is moved back and forth 
unphysically. This problems are avoided when TCAF solution is chosen. 
The complete spectral variation of the type seen in Cyg X-1 and other binaries are presented in C97. 

\begin{figure}
\vbox{
\vskip 0.0cm
\hskip 0.0cm
\centerline{
\psfig{figure=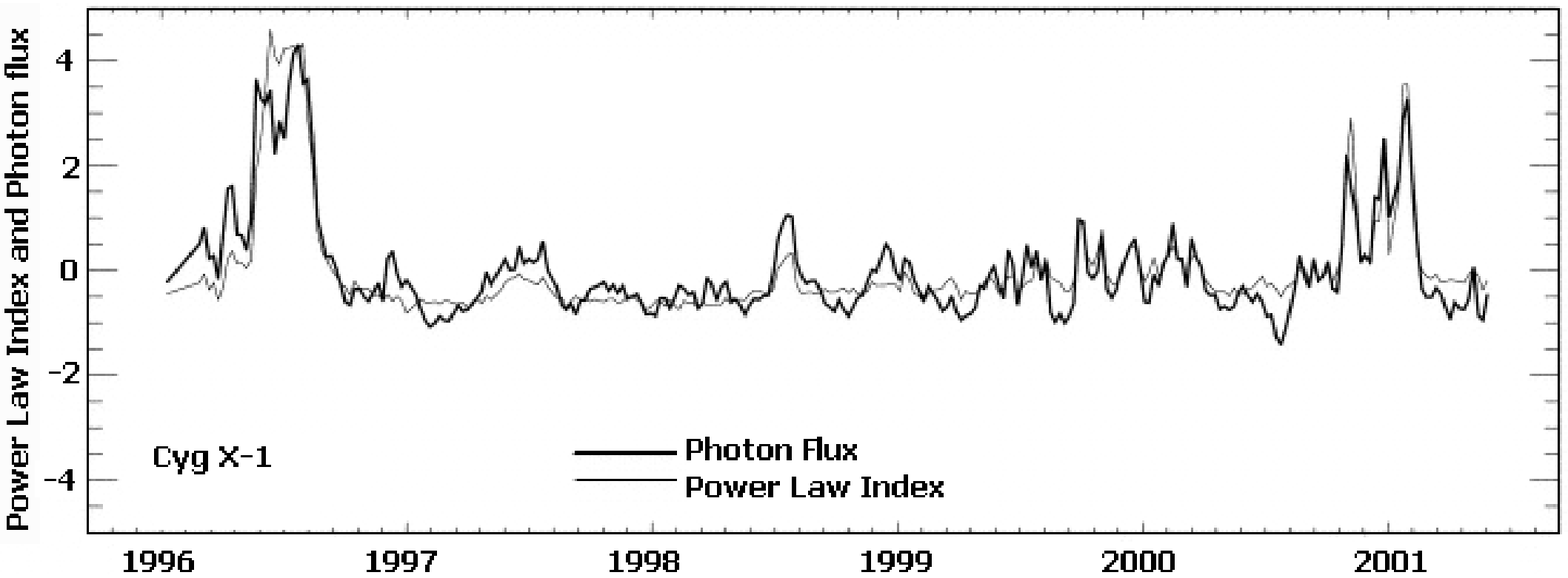,height=5truecm,width=11truecm}}
\vskip 0.1cm
\centerline{
\psfig{figure=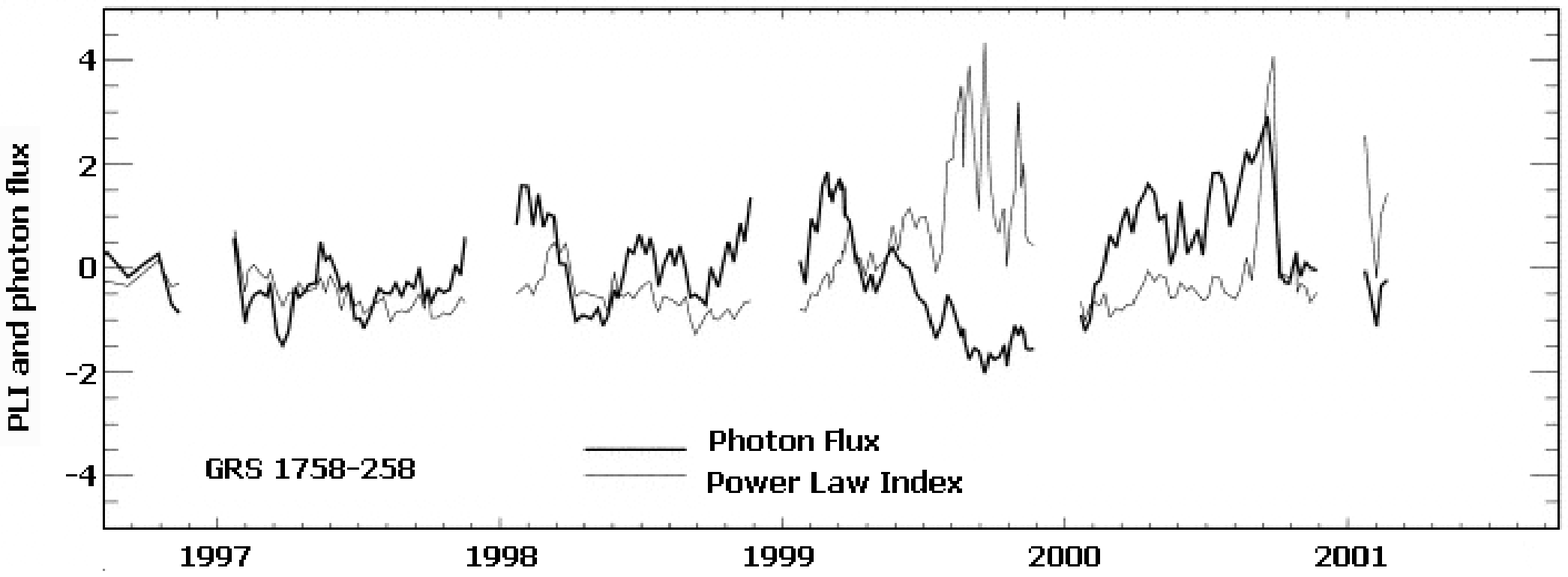,height=5truecm,width=11truecm}}}
\vspace{0.0cm}
\caption[]{ ASM light curves for a wind accretor (Cyg X-1) and a low-mass binary system (GRS1758-258) from RXTE 
data (Smith, Heindl and Swank, 2002) showing distinctly different behaviour due to the absence and presence
of a dominant Keplerian disk in these objects.}
\end{figure}

Having thus established that the behaviour of a large population of galactic black hole candidates can be 
understood only with TCAF solution, attempts were made to explain fast variabilities.
In case the shock oscillates back and forth (Molteni, Sponholz \& Chakrabarti, 1996; 
Molteni, Ryu, \& Chakrabarti, 1996; 
Ryu, Chakrabarti \& Molteni, 1997; Chakrabarti, Acharyya, Molteni, 2004),
the hard X-ray emission is likely to be modulated at the shock oscillation frequency. With 
the increasing of the Keplerian rate, i.e., cooling of the CENBOL, the shock moves
in due to fall of pressure and the frequency of QPO rises. This behaviour is also well known. 
According to the TCAF solution, different length scales of the flow are responsible for different 
QPO frequency range or the break frequency. For instance,  the  oscillation at the
transition radius is responsible for $0.1-0.3$Hz QPO, the oscillation of the shock is responsible for
$1-15$Hz QPO, the oscillation of the inner transient shock is responsible for the $60-450$Hz QPO and the 
cooling time scale of the sonic sphere of the jet is responsible for the $0.01-0.03$Hz QPO. The jets and outflows are 
also found to be coming out from the CENBOL region (Chakrabarti, 1999) and the outflow rates 
were computed exactly (Chakrabarti, 1999; Das and Chakrabarti, 1999) as a function of the 
inflow rate. These outflows are found to change the spectral slope of the outgoing radiation
indirectly since they modify the particle density at CENBOL. Since the presence of CENBOL is a 
function of the spectral state, this paradigm naturally explains the relation between the 
outflow rate and the size of the Comptonizing region. These behaviours have now been verified
by observers and model builders (e.g., Fender et al. 2000 and references therein). 

When the magnetic field is included, the synchrotron photons must be included into the solution
as they play a major role in cooling the sub-Keplerian flow. As the sub-Keplerian rate is 
increased, the density of soft photons goes up, but it needs not be sufficient to cool the 
electrons as the number of electrons themselves also goes up. 
Furthermore, the shocks in the flow can accelerate electrons to a high 
energy which in turn may be able to produce photons at very high energy $\gamma$-ray. 
In the next section, we shall discuss the spectral properties of black hole candidates in high energies.

\section{Spectral properties of galactic black holes in $\gamma$-rays}

The TCAF solution usually has a single static, oscillating or a propagating shock. However, numerical simulations 
suggest the presence of a transient shock just before the flow passes through the inner sonic point
(Samanta, Ryu, Chakrabarti 2004). Thus, there are possibilities to produce power-law high 
energy emissions by the power-law electrons generated by repeated passages through 
these shocks. Details are presented in Mandal \& Chakrabarti (2004) and some of the 
highlights are presented in Mandal \& Chakrabarti (this volume). 
It is entirely possible that the bulk of the high energy gamma-rays could be contributed 
by emissions from shocked regions of the outflow and jet. The physics of the shock formation 
in jets is not yet understood properly. Numerical simulations indicate periodic flaring of matter
emerging as outflows as the shock oscillated back and forth (Chakrabarti, Acharyya 
and Molteni, 2004). Figure 3 shows the oscillation of the inner region of the 
advective flow. Arrows indicate puffs of matter getting out in the alternating sides
during oscillation. Such oscillations of the accretion shocks are not only responsible for the 
QPOs, they may also be responsible for injecting non-linear perturbations in the 
outflows. These shocks may steepen up as they 
propagate away from the central body (see, Fig. 1) and cause acceleration of particles and 
synchrotron radiation. This is to be investigated in future. It is possible that other
triggering mechanisms, such as reconnection of magnetic fields near the base of the jet 
suddenly releases energy in the outflows producing shocks as well.

Figs. 4(a-b) shows generic nature of the spectra of galactic black hole candidates 
in both high and low states for two sets of black holes: Set A includes GROJ1719-24, 
GROJ0422+32 and CYGNUS X-1 while Set B includes GRO1655-40 and GRS 1915+105. The 
high energy regions are adapted from the results from McConnell et al. (2002), 
Ling \& Wheaton (2003, 2004), Case et al. (2004), etc. Here `high' and `low' states mean  
the soft and hard states using the nomenclature as in Chakrabarti \& Titarchuk (1995) (see also,
Tanaka \& Lewin, 1995) where the high power is observed in soft ($\sim 1$keV) and hard 
($\sim 40-100$)keV  energies, respectively. These states can also be defined using the energy spectral
index  $F_E \sim E^{-\alpha}$) in the $2-20$keV range: $\alpha >1$ for soft state 
and $\alpha <1$ for hard state. CT95 model has been discussed in the previous section.

\begin{figure}
\vbox{
\vskip -3.0cm
\hskip 1.0cm
\centerline{
\psfig{figure=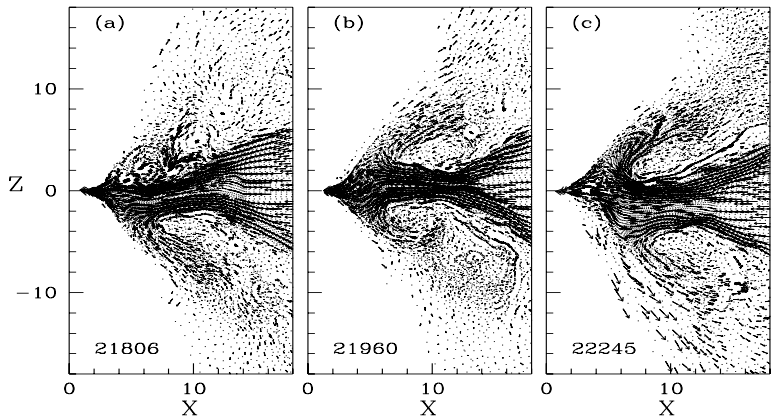,height=10truecm,width=13truecm}}}
\vspace{0.0cm}
\caption[]{Oscillating accretion shocks which are responsible for QPOs are also responsible for
variable outflows which steepen into shocks in the jets (Chakrabarti, Acharyya \& Molteni, 2004). }
\end{figure}

In the gamma-ray regime, the nomenclature is not unique yet. Ling \& Wheaton (2004)
like to call the above mentioned `canonical' high and low states 
as the `low $\gamma$ intensity state' and `high $\gamma$ intensity state' respectively. 
When seen in high energy gamma rays $\gsim 400$keV, the high state 
also corresponds to the high (hard) $\gamma$-ray state and similarly for hard states.
The difference in these two Sets (marked A and B) in Fig. 4(a-b) lies in the very high energy gamma-ray region: 
while in Set A, the so-called `high $\gamma$-ray intensity state' has a thermal spectrum 
below $200-300$keV and a weaker and softer power-law tail for higher energies, in Set B, the same
state is manifested as simply a continuous non-thermal power-law with or without a break (Case et al. 2004). 

\begin{figure}
\vbox{
\vskip 0.0cm
\hskip 0.0cm
\centerline{
\psfig{figure=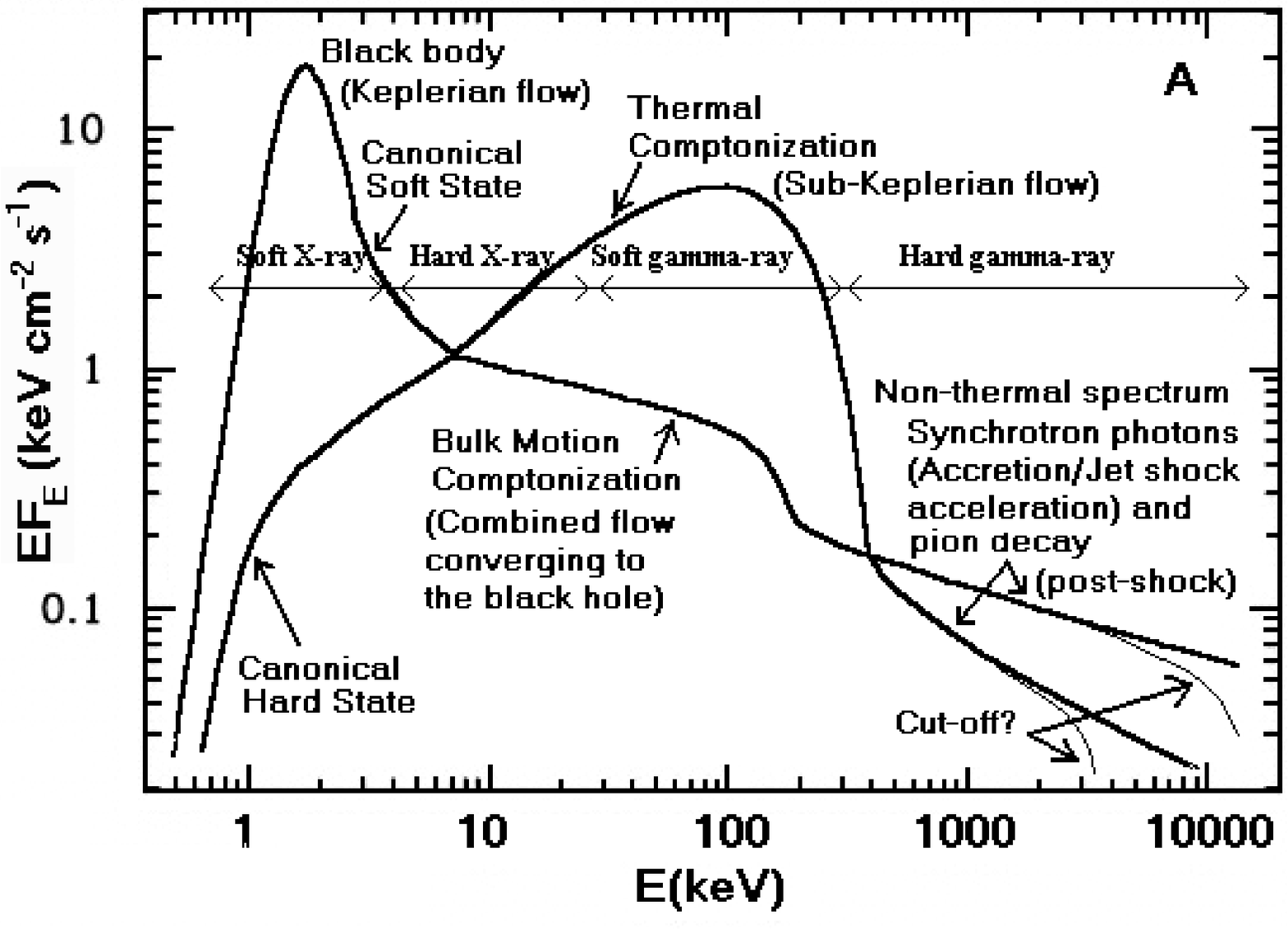,height=5truecm,width=8truecm}}
\vspace {0.5cm}
\hskip 2.0cm
\psfig{figure=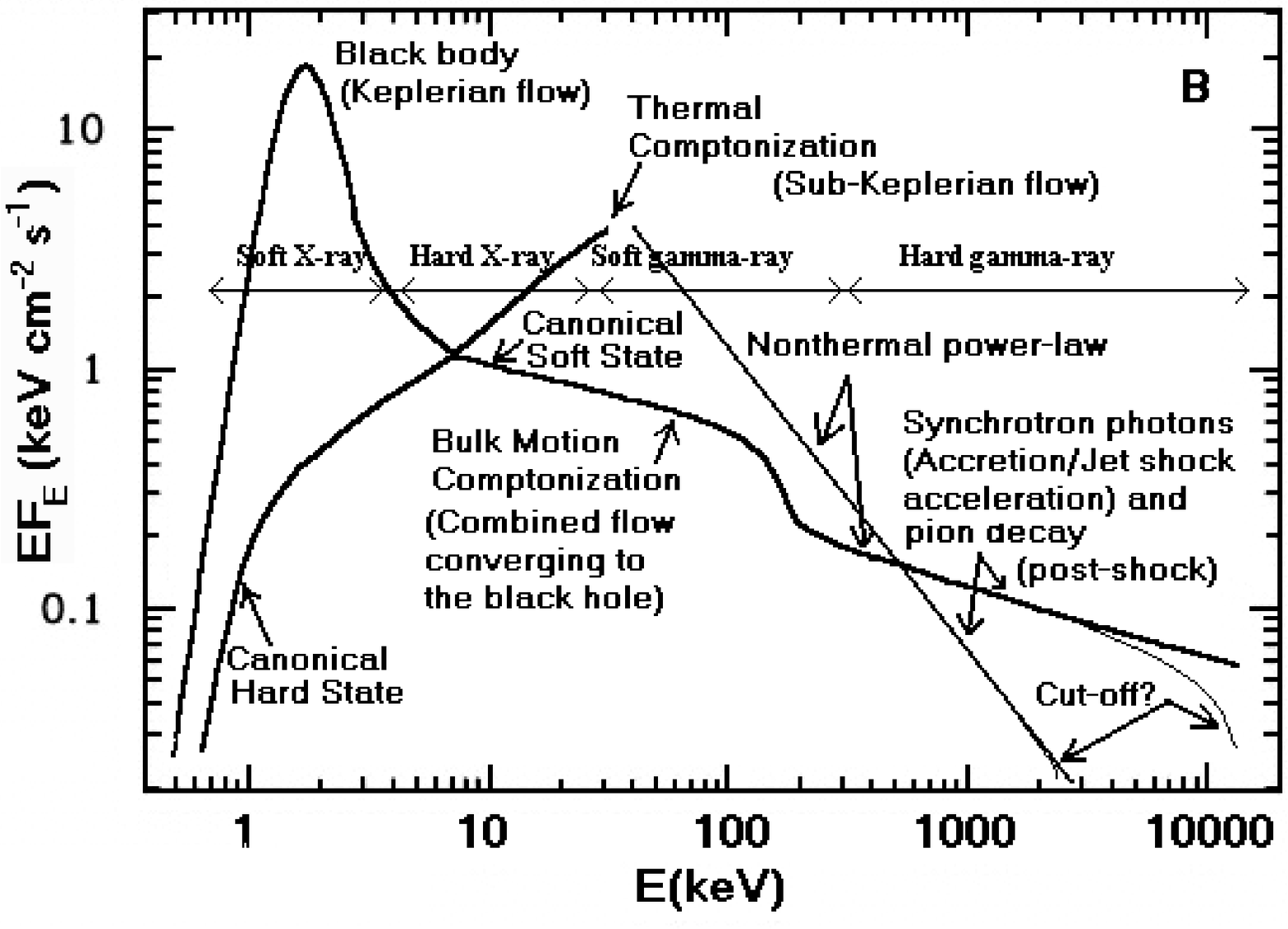,height=5truecm,width=8truecm}}
\vspace{-0.5cm}
\caption[]{Soft and hard state transitions of two classes of black holes which supposedly
differ in the very high gamma-ray region for the high (soft) gamma intensity state. In A,
it consists of thermal Comptonization till $200-300$keV and power-law after that. In B,
it consists of non-thermal power-law. }
\end{figure}

\begin{figure}
\vbox{
\vskip 0.0cm
\hskip 0.0cm
\centerline{
\psfig{figure=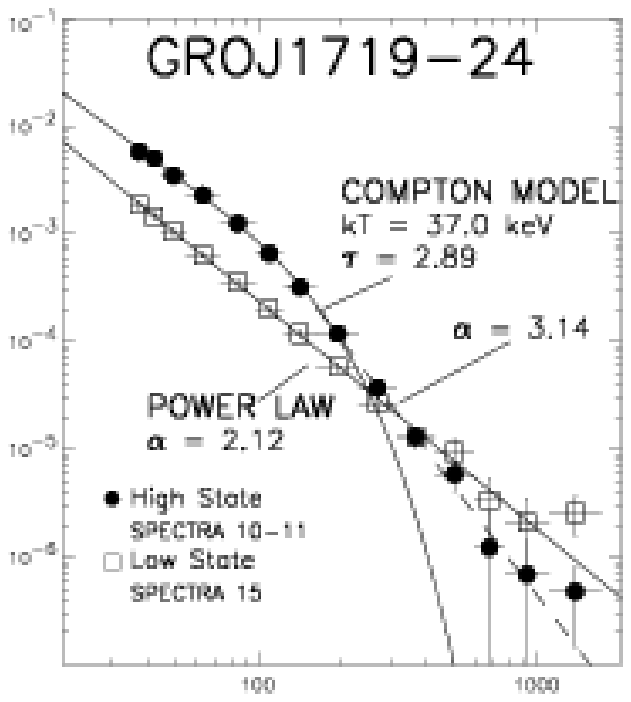,height=3.9truecm,width=3.9truecm}
\psfig{figure=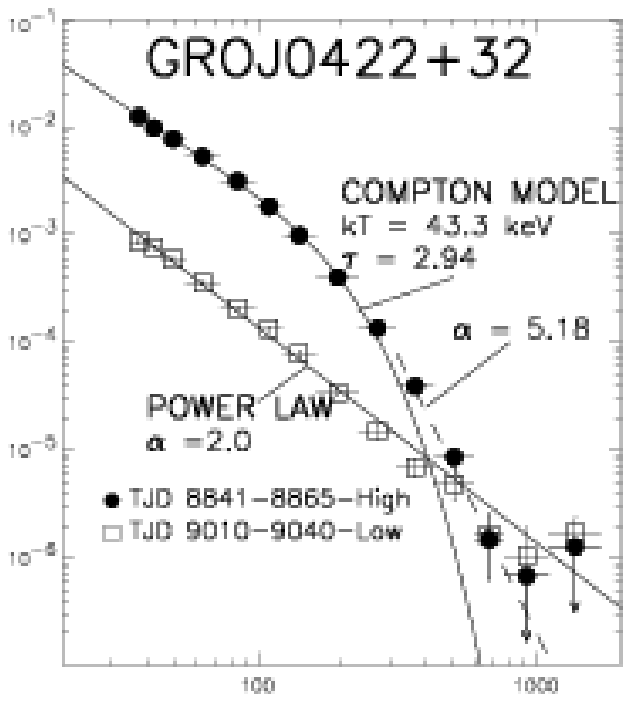,height=4truecm,width=4truecm}
\psfig{figure=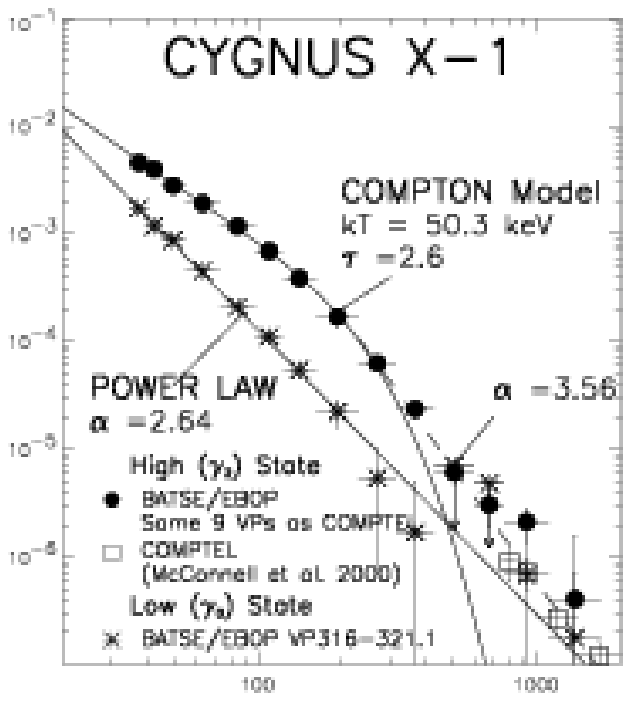,height=3.9truecm,width=3.9truecm}}}
\vspace{0.3cm}
\caption[]{Photon fluxes (in photons/cm$^2$/s/keV of three black hole candidates which show thermal 
Comptonization and a non-thermal power components in canonical hard states (from Ling et al. 2004).}
\end{figure}

\begin{figure}
\vbox{
\vskip -0.3cm
\hskip 0.0cm
\centerline{
\psfig{figure=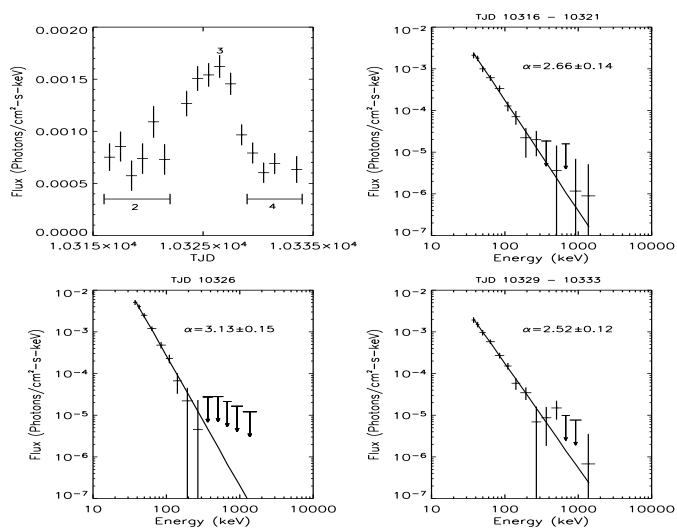,height=7truecm,width=9truecm}}}
\vspace{0.0cm}
\caption[]{ Example of non-thermal power-law in the very high energy gamma-ray region 
in the canonical hard state of GRO J1655-40 (from Case et al. 2004). Here, a flare
is analyzed. The 2nd, 3rd and the 4th panels refer to the spectra of the marked regions of the photon flux
shown in Panel 1. At the peak of the flare, the spectrum is steeper.
}
\end{figure}

Examples of the observations in the two sets are presented in Fig. 5 (from Ling and Wheaton, 2004) and 
in Fig. 6 (from Case et al. 2004). The latter Figures show the nature of changes which are taking place
in the power law in the high (soft) gamma-intensity state during a hard X-ray flare. The spectrum steepens
at the peak of the flare. Case et al. (2004) find that while Cyg X-1, GRO J0422+32, GRO J1719-24
belong to Set A (Fig. 4a), GRO 1655-40 and GRS 1915+105 belong to the Set B (Fig. 4b). It is interesting to note that
while both candidates in Set B are considered to be LMXRBs, the candidates in Set A are mixed: only Cyg X-1
is HMXRB while others are LMXRBs. Thus, this difference in the spectra is puzzling.
It is also important that for all the candidates in Set A,
the time lag in hard X-ray  with respect to the softer X-ray is around $0.1-1$s (van der Hooft 
et al. 1999), which point to an extended region of hard X-ray emission region. It is possible 
that these two properties of Set A candidates are related. The explanation,  may not  thus lie
in the difference in the primary composition of the accretion flow which is known to be predominantly of low 
angular momentum for HMXRBs and Keplerian for LMXRBs but to a common cause, namely, the shock formation.

Physically, it is clear that the accretion shocks and the shocks in jets and outflows could be responsible for the 
production of very high energy gamma rays. One of the advantages of the TCAF solution is that the accretion shocks,
jets and shocks in jets are built into the solution. Thus, one could imagine having shock acceleration of electrons
in the accretion and the power-law electrons producing power-law photons most naturally without having to 
guess about the origin of the shocks. As we mentioned, shocks in jets could be produced due to perturbations
of the CENBOL (e.g., oscillation of shocks) or sudden deposition of energy at the base of the jet
through reconnection of magnetic fields. Usual photopion production in the jets and gamma-ray production in pion
decay cannot be ruled out also. 

\begin{figure}
\vbox{
\vskip 0.0cm
\hskip 0.0cm
\centerline{
\psfig{figure=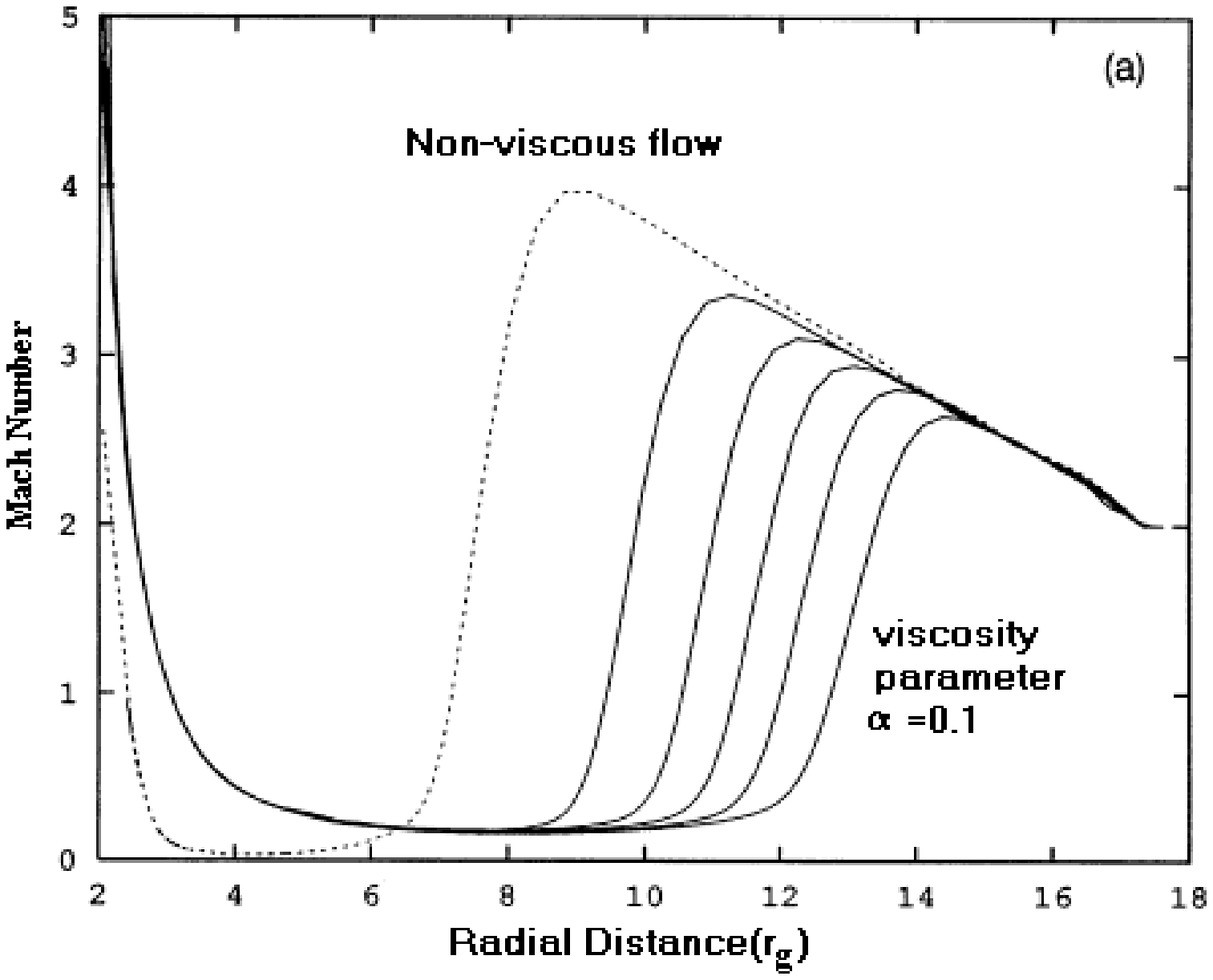,height=5truecm,width=8truecm}}
\vskip 0.1cm
\hskip 1.5cm
\psfig{figure=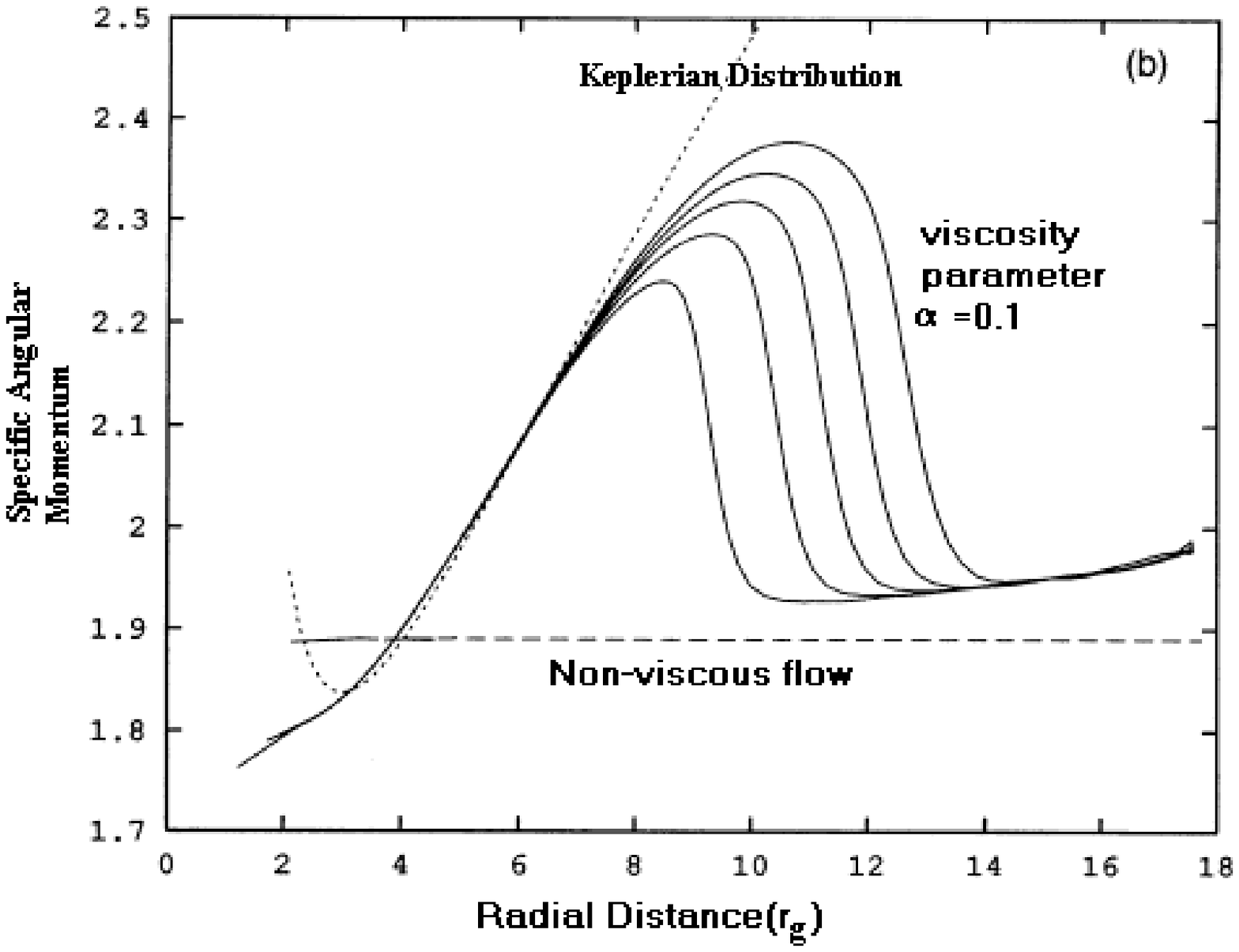,height=5truecm,width=8truecm}}
\vspace{0.0cm}
\caption[]{ Example of propagating shock solution in presence of larger viscosity
(Chakrabarti \& Molteni, 1995). Variation of (a) Mach number and (b) specific angular momentum as a 
function of the radial distance. Solutions for inviscid flow are also placed for comparison.}
\end{figure}

As a general comment, we might add that in the canonical soft states, the highest energy  gamma ray is
also the most intense (see Figs. 4(a-b)). One could ask the following questions:

\noindent i) If the highest energy $\gamma$-rays are coming from the jets, and at the same time, if the 
Keplerian rate is so high that the object is in the canonical soft state, then, can the jets or outflows  be 
very strong as well?

\noindent ii) If the jet and outflows are produced in harder states as the analytical 
work (Chakrabarti 1999) and some observations tend to show, then why the highest energy 
gamma rays are not seen in canonical hard states as well?

To understand this, one has to perhaps go back to the basics: Analytically it is easy to show that
the {\it ratio} $R$ of the outflow rate (${\dot M}_{out}$) to the inflow  rate (${\dot M}_{in}$) 
is very low in the soft state, and low in the hard state (see, Chakrabarti, 1999), but it is the
highest when the strength of the shock is intermediate. In other words, if the inflow rate is 
high, as in a canonical soft state, the outflow rate could also be high, though it needs not 
be relativistic.

\begin{figure}
\vbox{
\vskip 0.0cm
\hskip 0.0cm
\centerline{
\psfig{figure=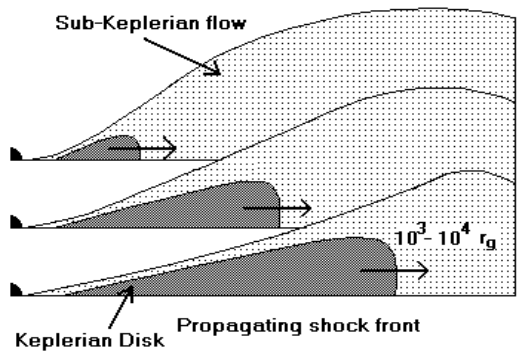,height=5truecm,width=7truecm}}}
\vspace{0.0cm}
\caption[]{Possible cause of soft state in Cyg X-1 and similar black hole candidates based on Chakrabarti
\& Molteni (1995) solution. In this case the inner part becomes Keplerian while outer region remains 
sub-Keplerian.}
\end{figure}

In the case of HMXRBs, such as Cyg X-1, the accretion is dominated by low-angular momentum
winds and it needs not produce any strong accretion shocks or jets as the centrifugal barrier 
needs not be strong enough even in the hard state. Indeed, broad QPOs at very low frequency 
($\sim 0.04$Hz) have been observed so far (Paul, Agrawal \& Rao, 1998) which indicates a 
shock at around $10^3-10^4r_g$. The jets are also found to be very slow, continuous
(Fender et al. 2000). So, it may be easy to understand why in canonical hard 
states, the very high energy gamma-ray is weaker.  In order to understand its behaviour 
in soft states, it may be pertinent to invoke the most relevant solution of this problem
presented by Chakrabarti \& Molteni (1995). Here, it was shown that the shock in an inviscid flow
tends to recede outward in presence of viscosity parameter higher than a critical value (Fig. 7a).
The post-shock subsonic region acquires a Keplerian distribution and the inner part essentially 
becomes a Keplerian disk (Fig. 7b). (In fact, this is how a Keplerian disk forms in the first place.) 
It is possible that the temporary presence of a large viscosity is the reason for Cyg X-1 to go 
to a soft-state. The sub-Keplerian region at $ r \sim 10^3 - 10^4 r_g$ could become the 
Comptonizing region (Fig. 8). It is not unlikely that some jets could come out from 
the interface of the Keplerian and sub-Keplerian flow taking away excess angular 
momentum. Could the shocks in this outer jet or at the propagating interface be responsible 
for the gamma-rays, as well as the time lag of $0.1-1$s as observed? In GRO 1655-40 and GRS 1915+105, 
jets are present and they also show various class transitions because of slow variation 
of accretion rates and because of interaction of jets with Comptonized photons. These 
cause blobby nature of the jets and perhaps stronger ultra-relativistic shocks
in jets which, in turn, produce shock acceleration of particles eventually producing 
power-law $\gamma$-rays as reported by Case et al. (2004). 

So far, we mostly discussed the solutions and interpretations in the case of galactic 
black holes. Fortunately, the nature of solutions does not change with the black 
hole mass and most of the considerations 
discussed above remain valid for supermassive black holes.
The major difference would be in timing properties as all the time intervals will 
scale with the central mass. Since in this case, we do not expect any `binary companion' 
to supply matter, the matter is likely to be from the winds of nearby stars or from 
disruption of the stars. Thus, the formation of steady, oscillating or propagating shocks
will be even more relevant.

\acknowledgements

The author acknowledges the organizers for the invitation. This work was made possible in part through a 
grant (Grant No. SP/S2/K-15/2001) from Department of Science and Technology (DST).

\end{article}

\begin{thebibliography}{}

\bibitem[]{}
Case, G.~L. et al., Observations of gamma-ray outbursts from galactic microquasars, {\em Chin. J. Astron. Astrophys.} (in press)

\bibitem[]{}
Chakrabarti, S.~K., Theory of transonic astrophysical flows, World Scientific Company, 1990.

\bibitem[]{}
Chakrabarti, S.~K., 
 Accretion disks in astrophysics, In  J. Franco et al. , editors, 
{\em Numerical Simulations in Astrophysics}, Mexico city, Mexico, 1993, Cambridge University Press, UK.

\bibitem[]{}
Chakrabarti, S.~K., 
Accretion disks in active Galaxies- the sub-Keplerian paradigm, In  H. B\"ohringer et al., editors,
{\em Proceedings of 17th Texas Symposium}, Munich, Germany, 1994, New York Academy of Sciences, New York.

\bibitem[]{} Chakrabarti, S.~K. \&  Molteni, D.  Viscosity prescriptions and the evolution of accretion disks, 
{\em M.N.R.A.S.}, 272, 80, 1995.

\bibitem[]{}
Chakrabarti, S.~K., 
Grand unification of solutions of accretions and winds around black holes and neutron stars, 
{\em Astrophys. J.}, 464, 664, 1996a.

\bibitem[]{}
Chakrabarti, S.~K., 
Accretion processes on black holes, {\em Phys. Rep.}, 266, 229, 1996b.

\bibitem[]{}
Chakrabarti, S.~K.
Estimation and effects of the mass outflow from shock compressed flow around compact objects, 
{\em Astron. \& Astrophys.},  351, 185, 1999.

\bibitem[]{}
Chakrabarti, S.~K. et al.,
The effect of cooling on time dependent behaviour of accretion flows around black holes,
{\em Astronom. \& Astrophys.}, 421, 1, 2004.

\bibitem[]{}
Chakrabarti, S.~K. \& Nandi A.,
Fundamental States of accretion/jet configuration and the black hole candidate GRS1915+105, {\em Ind. J. Phys.}  75(B), 1, 2000

\bibitem[]{}
Chakrabarti, S.~K. \& Titarchuk, L.G., 
Spectral properties of galactic and extragalactic black hole candidates, {\em Astrophys. J.}, 455, 623, 1995.

\bibitem[]{}
Das, T. K. \& Chakrabarti, S.~K.,
Computation of the mass outflow rate from neutron star and black hole accretion disks, 
{\it Classical and Quantum Gravity}, v. 16, No. 19, 3879, 1999.
 
\bibitem[]{}
Esin, A. A. et al., Spectral transitions in Cygnus X-1 and other black hole x-ray binaries, {\it Astrophys. J.}, 505, 854, 1998.

\bibitem[]{}
Fender, R. et al.,
Quenching of the radio jet during the X-ray high state of GX 339-4, {\it Astrophys. J.}, 519, L165, 1999.

\bibitem[]{}
Fender, R.P. et al. 
The very flat radio-millimetre spectrum of Cygnus X-1, {\em MNRAS} 312, 853, 2000.

\bibitem[]{}
Galeev, A.~A. et al., Structured coronae of accretion disks, {\em Astrophys. J.}, 229, 318, 1979.

\bibitem[]{}
Ling, J.~C. \&  Wheaton,  W.~A.,
BATSE soft gamma-ray observations of GRO J0422+32
{\em Astrophys. J.}, 584, 399, 2003.

\bibitem[]{}
Ling, J.~C. \&  Wheaton,  W.~A.,
Gamma-ray spectral characteristics of thermal and non-thermal emission from three black holes,
{\em Chin. J. Astron. Astrophys.} (in press)

\bibitem[]{}
Mandal, S. \& Chakrabarti, S.~K.
Signature of accretion shocks
in emitted radiation from a two temperature advective flows around black holes, {\em Astron. \& Astrophys.}, (submitted)

\bibitem[]{}
McConnell, M.~L.  et al.,
The soft gamma-ray spectral variability of Cygnus X-1, {\it Astrophys. J.}, 572, 984, 1998.

\bibitem[]{}
Molteni, D. et al., Resonance oscillation of radiative shock waves in accretion disks around
a black hole, {\it Astrophys. J.} 457, 805, 1996.

\bibitem[]{}
Nowak, M., State transitions in  black hole candidates, In P. Durouchoux et al., editors,
New Views on Microquasars, Corsica, 2002, Centre for Space Physics, Kolkata, India.

\bibitem[]{}
Paul, B. et al.,
Low frequency quasi-periodic oscillations in the hard X-ray emission from Cygnus X-1,
{\em J. Astrophys. Astron.}, 19, 55, 1998.

\bibitem[]{}
Ryu, D. et al., Zero Energy
Rotating accretion flows near a black hole, {\it Astrophys. J.}, 474, 378, 2004.

\bibitem[]{}
Samanta, M.~M. et al.
Numerical simulation of double shock structure in accretion flows around black holes (submitted)

\bibitem[]{}
Smith, D. M. et al.,
Two different long-term behaviours in black hole candidates: evidence for two accretion
flows?, {\em Astrophys. J.}, 569, 362, 2002.

\bibitem[]{}
van der Hooft, F. et al.,
Hard x-ray lags in GRO J1719-24,  {\it Astrophys. J.}, 519, 332, 1999.

\bibitem[]{}
Zdziarski, A.A. \& Gierlinski, M., Radiative processes, spectral states and variability of
black hole binaries, {\em Prog. Theo.Phys}, (in press).

\end{thebibliography}
\end{document}